\documentclass[useAMS,usenatbib]{mn2e}

\usepackage{graphicx}
\usepackage{deluxetable}
\usepackage{longtable}
\usepackage[varg]{txfonts}

\title[H$\alpha$ photometry in 47 Tucanae]{H$\alpha$ photometry of low mass stars in 47 Tucanae: 
chromospheric activity and exotica}
\author[]{
G. Beccari$^{1}$\thanks{E-mail: gbeccari@eso.org} 
G. De Marchi$^{2}$
N. Panagia $^{3,4,5}$ and
L. Pasquini$^{1}$
\\
$^{1}$European Southern Observatory, Karl--Schwarzschild-Strasse 2,
85748  Garching bei M\"unchen, Germany\\
$^{2}$ESA, Space Science Department, Keplerlaan 1, 2200 AG Noordwijk, The
Netherlands\\
$^{3}$Space Telescope Science Institute, Baltimore, MD 21218, USA\\
$^{4}$INAF-CT, Osservatorio Astrofisico di Catania, Via S. Sofia 78, 95123
Catania, Italy\\
$^{5}$Supernova Ltd, OYV \#131, Northsound Road, Virgin Gorda, British Virgin
Islands}
\begin{document}

\date{}

\pagerange{\pageref{firstpage}--\pageref{lastpage}} \pubyear{2002}

\maketitle

\label{firstpage}

\begin{abstract}
We have used archival {\it Hubble Space Telescope} observations obtained
with the {\it Advanced Camera for Surveys} to study the H$\alpha$
emission properties of main sequence stars in the globular cluster 47
Tucanae. Using a combination of multi-band observations in the F606W,
F814W and F658N bands,  we search for stars showing H$\alpha$ excess
emission. An accurate photometric measurement of their H$\alpha$
equivalent width allows us to identify objects with large H$\alpha$
emission, which we attribute to mass  accretion rather than enhanced
chromospheric activity. The spatial position of some of these stars is
coincident with that of known X-ray sources and their location in the
colour-magnitude diagram allows us to classify them as active binaries
or cataclysmic variables (CVs). We show that this method, commonly 
adopted to study accreting discs in young stellar objects,    can be
successfully used to identify and characterise candidate CVs.

\end{abstract}


\section{Introduction}

Globular Clusters (GCs) are very efficient ``kilns'' for forming exotic
objects, such as low-mass X-ray  binaries (LMXBs), cataclysmic variables
(CVs), millisecond pulsars (MSPs), and blue straggler  stars (BSSs).
Most of these exotica result from the evolution of binary systems
originated and/or hardened by stellar interactions in the dense cluster
cores, thus serving as a diagnostic of the dynamical evolution of
GCs~\citep[][]{ba95}.  In particular, CVs are semi-detached binary
stars, consisting of a white dwarf (WD) primary accreting from  a
main-sequence (MS) or a sub-giant companion~\citep[see][for a recent
review]{kn11}  while LMXBs and MSPs contain a neutron star (NS).

The accretion process leaves some characteristic signatures in the
spectrum of these stellar exotica. For instance, a typical feature
observed in the spectra of CVs is emission from Balmer recombination
lines~\citep[e.g. H$\alpha$, H$\beta$; see][]{wit06} or excess in the
ultraviolet (UV) continuum~\citep[see e.g.][]{fe01}. ~\citet[][]{wil83}
suggest that an accretion disc is the dominant source of optical and UV
emission, whereas line emission may originate from the outer edge of the
disc or the accretion corona. In magnetic CVs, optical and UV emission
likely  generate from accreted material that flows along magnetic field
lines, and accretes directly onto the WD at/near  its magnetic poles
without forming a viscous disc. However, UV and line emission also
originates from various parts of an accretion disc in non-magnetic or
only weakly magnetic CVs~\citep[see e.g.][]{wit06}.

A different source of emission lines in the spectra of MS
stars is chromospheric activity (CA), due to processes that make the
temperature of the outer atmosphere higher than it would be if radiative
equilibrium held. Such non-radiative heating mechanisms are powered by
convection and magnetic fields.  In the outer layers of these stars, the
temperature is increasing towards the surface, and  the main cooling
mechanism is radiative loss through strong resonance lines, such as  Ca
II H and K, Mg II h and k, H$\alpha$~\citep[][]{ly80}. Signatures
of H$\alpha$ emission from CA in spectra of low mass stars are thus
expected and a detailed knowledge of the H$\alpha$ equivalent width
(hereafter EW(H$\alpha$)) produced by CA is crucial when searching for
H$\alpha$ emission signatures from accreting stars of all ages. 

A number of studies have looked for accreting binaries through emission
lines searches in GCs~\citep[e.g.][]{po02,kon06,ba08,coh10,lu11}. The GC
47 Tucanae is definitely one of the most promising targets for this
kind of investigation, since it harbours a crowded zoo of stellar
exotica, including BSSs, LMXBs and candidate CVs~\citep[e.g.][]{fe01}. 
~\citet[][hereafter H05]{he05} published a
catalogue of 300 X-ray sources inside the half-mass radius of 47
Tuc~\citep[r$_h=3\farcm17$;][]{ha96} obtained from deep X-ray
observations with the {\em Chandra X-ray Observatory}. Many of these stars
are classified as CVs, interactive binaries, quiescent LMXBs (qLMXBs) and MSPs, and
therefore a signature of H$\alpha$ excess emission should be expected
for at least some of them. Recently~\citet[][]{kn08} presented far-UV
spectroscopic observations obtained with the {\em Hubble Space
Telescope} (HST) of 48 blue objects  in the core of 47 Tuc. These
authors provide spectroscopic confirmation of 3 CVs via the detection of
emission lines and find new evidence for dwarf nova eruptions from two
of them. However, a large number of objects in the H05 catalogue still
remain unclassified.

Following a method presented by~\citet[][hereafter DM10]{de10}
to detect and study stars with H$\alpha$ excess emission in star forming
regions, we have used archival HST observations acquired with the {\it
Advanced Camera for Surveys} (ACS) to perform an H$\alpha$ survey of MS
stars in 47 Tuc. 
The method of DM10 uses a combination of V and I  broad-band photometry
with narrow-band H$\alpha$ imaging to measure the EW(H$\alpha$)
of stars showing H$\alpha$ excess emission, thereby offering a new ad very efficient
way to identify and characterise accreting stars in GCs. In this work
we present the method and we explore the possibilities offered by its 
application to 47Tuc.

In Section~\ref{sec_obs} we describe the dataset and the reduction
strategy while the DM10 method is described in Section~\ref{sec_ha}. We
applied the method of DM10 with two main goals. Firstly, we searched for
H$\alpha$ excess emitters amongst the cluster stars located at a radius
$r>70\arcsec$, where the photometric completeness allows us to perform
a homogeneous H$\alpha$ survey of all the spectral types sampled by
our data (Section~\ref{sec_ch}). Secondly (see Section~\ref{sec_x}), we 
searched for optical counterparts to the H05 X-ray sources by looking for
objects with H$\alpha$ excess emission inside the X-ray error circle,
regardless of their location in the cluster. In Section~\ref{sec_cmd} we
conclude by discussing the possible nature of some unclassified objects
from their location in the colour-magnitude diagram (CMD) of the
cluster. 


\section{Observations and data reduction}
\label{sec_obs}

\begin{table}
\caption{Log of the observations}             
\label{tab_obs}      
\centering                          
\begin{tabular}{c c c c c}        
\hline\hline                 
Filter & N. of exposures & Tot. exposure time  & Prop. ID & Date of obs.\\    
\hline                        
   F606W & 4 & 200s & 10775 & 13.03.2006\\      
   F814W & 4 & 200s  & 10775 &--\\
   F658N & 7  &  2630s & 9281 & 30.09.2002\\
   F658N & 7  &  2630s & 9281 & 02.10.2002\\
   F658N & 6  &  2180s & 9281 & 11.10.2002\\
\hline                                   
\end{tabular}
\end{table}

\begin{figure}
\centering
\includegraphics[width=9cm]{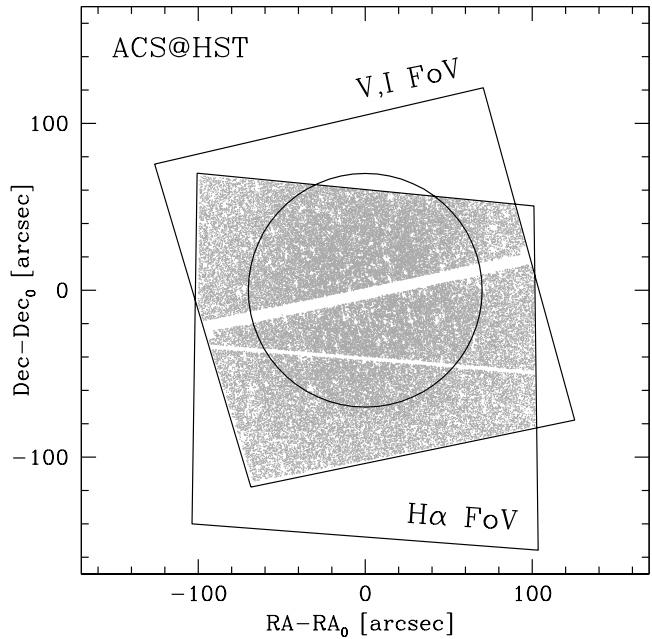}
\caption{Map of the combined ACS observations. The position of the stars
are plotted with respect to the centre of gravity from~\citet[][]{fe04}. The images in the
inner $70\arcsec$ from the cluster centre (circle) suffer from strong
crowding. }


\label{fig_fov}
\end{figure}

The data used in this work consist of a series of archival HST/ACS images
obtained through the broad F606W and F814W bands and with the
narrow-band F658N filter (hereafter V, I and H$\alpha$, respectively).
The images, acquired with the ACS in the Wide Field Channel (WFC) mode,
cover a total field of view (FoV) of  $\sim3\farcm4\times3\farcm4$ with
a plate-scale of $0\farcs05$ per pixel. The V- and I-band observations
come from proposal 10775 while the H$\alpha$ band data belong to
the HST proposal 9281. Details on exposure time
for each filter are given in Table~\ref{tab_obs}. The images sample the
cluster centre, but slightly different pointings and telescope
orientations have been used to perform these observations. As a
consequence, the sampled area in common between the two datasets is
$\sim40\,\%$ less than the whole FoV. The geometry of the FoV of 
the combined datasets is shown in Figure~\ref{fig_fov}. 

\subsection{Photometry of the V and I band data set}
The V- and I-band images were analysed using the standard 
DAOPHOTII~\citep{ste87} point spread function (PSF) photometric
reduction procedure. More than 50 isolated and well sampled stars were
used on each flat-fielded (FLT) image\footnote{The FLT images were
corrected for geometric distortions and for the effective area of each
pixel following the approach described in~\citet[][]{sir05}. Correction for losses
of Charge Transfer Efficiency was performed on the FLT images following~\citet[][]{an10}} to
accurately calculate a PSF model. PSF fitting was performed on every
image with the DAOPHOTII/ALLSTAR routine and a master list was obtained
which contains all objects that have been detected in at least three of
the four individual exposures in each of the V and I bands. The
instrumental magnitudes were then calibrated and  brought to the VEGAMAG
photometric system using the procedure described in~\citet[][]{sir05}.
A sample of bright isolated stars has been used to transform the instrumental 
magnitudes to a fixed aperture of $0\farcs5$, and the extrapolation to infinite 
and transformation into the VEGAMAG photometric system was performed 
using the updated values listed in Tables 5 and 10 of~\citet[][]{sir05}\footnote{The new values are
available at the STScI web pages: http://www.stsci.edu/hst/acs/analysis/zeropoints}.

\begin{figure}
\centering
\includegraphics[width=9cm]{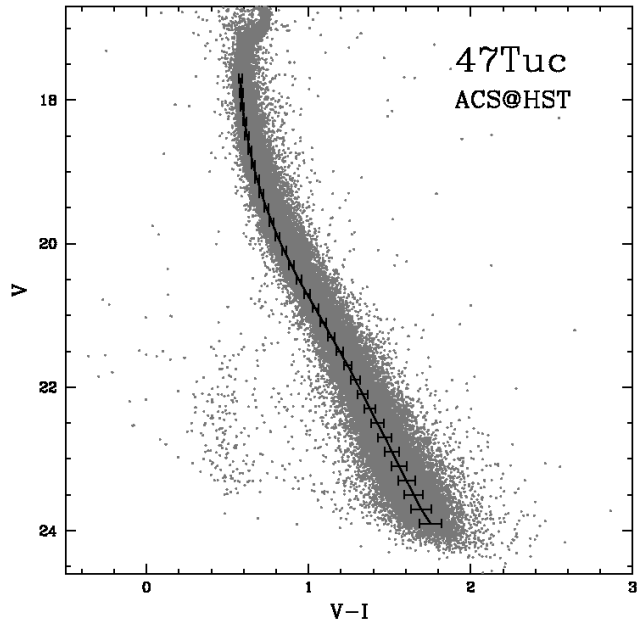}
\caption{CMD of the $\sim100000$ stars sampled in the V and I  data-set
of 47tuc. The MS ridge-line is shown as a solid line, together with
the average deviation (horizontal error bars in the plot). The solid line is not 
centred because of binaries and blends.}
\label{fig_cmd_raw}
\end{figure}

The final catalogue contains about 100\,000 stars,
sampling the MS stars down to $6.5$ mag below the turn off (TO). The CMD of
the stars detected in the V and I bands is shown in Figure~\ref{fig_cmd_raw}. Cluster stars 
are shown as grey points, while the location of the MS mean ridge line obtained with a second order 
polynomial interpolation with a 2.5 sigma rejection, is displayed as a black solid line together with
the average deviation (horizontal error bars in the plot).
Finally, we used $\sim3000$ stars in common with a  ground based catalogue obtained
with the Wide Field Imager at the MPE-2.2m telescope, published by~\citet[][]{fe04} to derive an 
astrometric solution and obtain the absolute right ascension (RA) and declination (Dec) 
positions of the stars sampled in the ACS catalogue. The astrometric coordinates of the stars in the WFI catalogue were registered with
the ICRS reference system adopting a GCS2.2 catalogue of astrometric standards. The r.m.s. scatter of the final 
solution was $\sim0\farcs3$ in both RA and Dec.

\subsection{Photometry of the H$\alpha$ band data set}
A different approach was used for the H$\alpha$ dataset.  Three drizzled
(DRZ) images, generated by the standard pre-reduction pipeline 
using the IRAF {\em multidrizzle} package, were
retrieved  from the HST archive. The DRZ images are a combination of 
7, 7 and 6 FLT images, for an effective total exposure time of 2630\,s,
2630\,s and 2180\,s, respectively (see also Table~\ref{tab_obs}).
Besides being cleaned from cosmic rays and detector imperfections, the 
DRZ images offer higher signal-to-noise ratio compared to the single
FLT exposures and therefore a more precise measure of the H$\alpha$
magnitudes.

The H$\alpha$ images are affected by very low crowding conditions. 
This allowed us to safely use aperture photometry to measure the stellar magnitudes in the 
H$\alpha$ band. Moreover, as documented in the MULTIDRIZZLE Handbook, the distortion present 
in the imaging instruments on board HST produces sampling patterns that are not uniform 
across the field, due to the changing pixel size. This may cause a ``blurring'' in the
PSF of the stars in the processed images and directly impacts the uniformity of the output PSF,
making the PSF modelling in DRZ images slightly more complicated than in the FLT images. 

We performed aperture photometry with the IRAF {\em phot} task on each
DRZ image using as input coordinates the positions of the stars in the 
master list obtained in the V and I bands, after registering the 
coordinate system of each DRZ image. In order to do that
a preliminary aperture photometry was performed on the DRZ images with the 
Sextractor photometric package~\citep[][]{be96} with the aim of producing
a list of centroid and magnitudes of the stars in the three H$\alpha$ frames. 
These lists of stars were used as reference to accurately transform the coordinates 
of all the stars in the V and I master list onto each DRZ frame position.  Cross-correlations of
the catalogues were performed with CataXcorr, a code developed and
maintained by Paolo Montegriffo at INAF-Bologna Astronomical Observatory~\citep[see e.g.][]{fe04}.
More than 20000 stars in common between the V and I master list and each
DRZ star lists from aperture photometry allowed us to reduce the uncertainties
(i.e. the X and Y r.m.s.) of the transformations to $\sim4\times10^{-6}$ pix.

The H$\alpha$ magnitudes of the
stars were measured inside a radius of 2 pixels, while the background
was locally estimated inside an annulus between 4 and 7 pixel radius
around each star. The photometric uncertainty is defined as the standard
deviation of three independent measurements of the same object. 
The H$\alpha$ instrumental magnitudes obtained in this way were finally
calibrated to the VEGAMAG photometric system following the procedure
of~\citet[][]{sir05}. 

\section{Selection of objects with H$\alpha$ excess emission}  
\label{sec_ha}

As  mentioned in the introduction, our first goal is to apply the method
developed by DM10 to perform a systematic and homogeneous search for
candidate H$\alpha$ excess emitters along the cluster MS. As mentioned in the
previous section, we have obtained H$\alpha$ magnitudes for all the objects
sampled in the $V$ and $I$ band images. However, crowding conditions vary 
considerably across the field covered by our observations and the homogenous 
search for stars with H$\alpha$ excess emission is harder to achieve in the most
central regions of the cluster. Therefore, in this first
part of the paper (Sections\, \ref{sec_ha} and \ref{sec_ch}) we study the stars located at a distance larger 
than $70\arcsec$ from the cluster centre (black circle in
Figure~\ref{fig_fov}),  a region which is not affected by stellar
crowding.  Stars with H$\alpha$ excess in the central cluster regions are discussed in
Section\,\ref{sec_x}. We refer to DM10~\citep[see also][]{spe12} for a detailed
description of the method, while here we illustrate the crucial steps
needed to successfully apply it. 

The first step consists in correcting the stellar magnitudes for
reddening, which we perform by assuming the canonical extinction value
towards 47\,Tuc, namely $E(B-V) = 0.04$~\citep[][]{ha96} using
the~\citet[][]{car89} extinction law with Rv=3.1. 
It is important to clarify here that, while reddening correction might be a
rather complex step when studying star-forming regions, where stars are
still embedded in the molecular cloud that may cause strong differential
extinction, this is not at all an issue in a GC where the stars are
subject to very little or no differential reddening. In particular,
giving the small amount of total reddening affecting the stars in 47\,Tuc, 
differential reddening is rather unlikely~\citep[see][]{an09}. Moreover, in~\citet[][]{be10} we
demonstrated that even  over/underestimating $E(V-I)$ by much as 0.2 mag,
by far larger than the uncertainty of our assumption on the reddening,
would not change significantly the selection of stars with H$\alpha$
excess emission.  

\begin{figure}
\centering
\includegraphics[width=10cm,height=11cm]{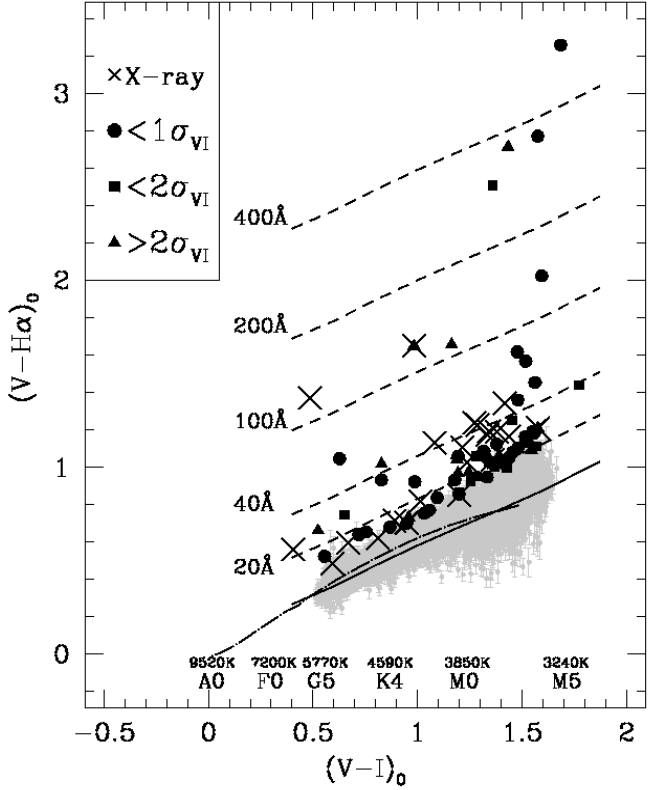}
\caption{Selection of stars with H$\alpha$ excess emission in a
colour--colour diagram. 
The solid line represents the median $(V-H\alpha)_0$ color of stars with an error 
on $(V-H\alpha)_0<0.1$, and is defined as the locus of stars without H$\alpha$ excess emission
and hence the location of stars with EW(H$\alpha$)=0. Dashed lines show the 
position of stars at increasing levels of H$\alpha$ emission. All the
objects with H$\alpha$ excess emission at least 5 times larger than the
photometric uncertainty on their $V-H\alpha$ colours are shown in
black. X-ray sources showing H$\alpha$ excess emission are shown
as crosses, while solid circles, squares and triangles indicate H$\alpha$ excess emitters
lying on the CMD at $(V-I)_0$ distance from the MS mean-ridge line, 1, 2 or more than 2 times the combined photometric 
uncertainty on the colour. The dot-dashed line shows the location of the colour relationship 
derived for these bands using the~\citet[][]{be98} atmospheric models.}
\label{fig_viha}
\end{figure}

Once the magnitudes of the stars are corrected for reddening, the
selection of the objects with H$\alpha$ excess emission is performed in
the $(V-H\alpha)_0$ vs. $(V-I)_0$ colour--colour diagram, as shown in
Figure ~\ref{fig_viha}. The traditional approach to search for sources
with H$\alpha$ excess is based on the use of the R-band magnitude as an
indicator of the level of the photospheric continuum near the H$\alpha$
line, so that stars with strong H$\alpha$ emission will have a large
$R-H\alpha$ colour~\citep[see e.g.][]{ba96,wit06,pa10}. However, as
discussed in DM10, since the R band is over an order of magnitude wider
than the H$\alpha$ filter, the $R-H\alpha$ colour does not provide 
a direct measurement of the H$\alpha$ equivalent width EW(H$\alpha$). 
Conversely, an accurate determination of the continuum is obtained using
the V, I and H$\alpha$ magnitudes combination shown in
Figure~\ref{fig_viha}, because the contribution of the H$\alpha$ line to
the V and I magnitudes  is completely negligible. Therefore the value of
the EW(H$\alpha$) can be directly derived. The $(V-I)_0$ colour is in
addition a useful colour index  for temperature determination and its
use allows one to take into account the variation  of the stellar
continuum below the H$\alpha$ line at different spectral types.  
The effective temperatures and spectral types shown in Figure~\ref{fig_viha}, are obtained 
adopting the atmospheric model of~\citet[][]{be98} and the specific ACS filters used in this 
investigation. As a reference, a star with an effective temperature of $\sim 4800$\,K has 
colour F555W$-$F814W$=1$ and F606W$-$F814W$=0.78$. The latter 
is the filter pair used in our photometry.

Note that, while the observations in the V and I bands are simultaneous,
those in the H${\alpha}$ band were collected at a different epoch.
Therefore, if there is variability in the continuum, the true V-H$\alpha$
values could be slightly different from those shown in
Figure~\ref{fig_viha}. However,~\citet[][]{ed03b} show that the magnitude
of the majority of stars in this field varies by $\sim 0.02$ mag. This
level of variability in the continuum does not affect our selection of
bona-fide H$\alpha$ excess objects.

The reference line (solid line in Figure\,\ref{fig_viha}) with respect
to which the H$\alpha$ excess emission is measured is empirically 
defined as the median $(V-H\alpha)_0$ colour of stars with combined
photometric errors smaller than $0.1$\,mag (grey points in Figure
~\ref{fig_viha}). Note that our empirical determination agrees very
well  with the colour relationship derived for these bands using the
\citet[][]{be98} atmospheric models (dot-dashed line in
Figure\,\ref{fig_viha}). Since the contribution of the H$\alpha$ line to
the $V$ magnitude is completely negligible, the H$\alpha$ excess
emission can be simply obtained as the distance $\Delta H\alpha$ of a
star from the empirical line, along the $(V-H\alpha)_0$ axis at the same
$(V-I)_0$ colour.  We consider as objects with bona-fide excess emission
only stars showing a $\Delta H\alpha$ at least value 5 times larger than
their combined photometric uncertainties in the V and H$\alpha$ bands
($\sigma_{VH\alpha}$). 

As already mentioned, the V and I deep images are strongly affected by
saturation of the bright giants, which produce diffraction spikes on the
images~\citep[e.g. see][]{an09}. This implies that the magnitude of a
star falling on the spikes cannot be trusted because of the uncertainty
in the background estimation. We visually inspected all the candidate
stars with H$\alpha$ emission and decided to conservatively discard a
star if a spike was falling inside the annulus around the stellar
centroid adopted for the local estimate of the background (7 pixels).

Once the $\Delta H\alpha$ value is calculated as explained above, the
equivalent width EW(H$\alpha$) for each star can be derived as 
$EW(H\alpha)=RW\times[1-10^{-0.4\times\Delta H\alpha}]$ (see Equation\,4
in DM10) where $RW$ is the rectangular width in $\AA$ of the filter, which
depends on the specific characteristics of the filter (see Table\,4 in
DM10).

A total of 59 stars outside of 70\arcsec from the cluster centre are
found to have H$\alpha$ excess emission. They are marked with black
symbols in Figure~\ref{fig_viha}, while the coordinates and magnitudes
of these objects, together with the value of EW(H$\alpha$), are listed in
Table~\ref{tab_emett}. Thick dots in Figure~\ref{fig_viha} denote
H$\alpha$-excess stars with a colour distance  $\Delta (V-I)_0$ from the
MS mean ridge line in Figure~\ref{fig_cmd_raw} equal to, or smaller
than, the photometric uncertainty  on their color  $\sigma_{VI}$
($\Delta(V-I)_0\leqq \sigma_{VI}$), squares denote H$\alpha$-excess 
stars with $\sigma_{VI}<\Delta(V-I)_0\leqq 2\sigma_{VI}$, and triangles
correspond to  H$\alpha$-excess stars with a colour distance
$>2\sigma_{VI}$. This is a simplified way to indicate the probability
that a star falls on the MS,  since the larger is the colour distance
the lower the probability of being a single MS star.  

The dashed lines in Figure\,\ref{fig_viha} correspond to EW(H$\alpha$)
thresholds of 20, 40, 100, 200 and 400\AA. Notice that we are assuming a
feature with positive EW as being in emission and negative EW as being
in absorption. As expected, the  majority of cluster stars show very low
or no H$\alpha$ emission.  Only 12 stars with spectral type between F0
and M0 show  EW(H$\alpha$)$>20\AA$. The number of objects with H$\alpha$
excess emission increases at later spectral types, but still only $\sim
0.2\,\%$  of the total number of objects sampled in this area show
excess emission with EW(H$\alpha$)$>20\AA$.  The distribution of these
stars in the cluster CMD will allow us to investigate more precisely the
nature of these objects (see Section~\ref{sec_cmd}).

\begin{figure} 
\centering
\includegraphics[width=9cm]{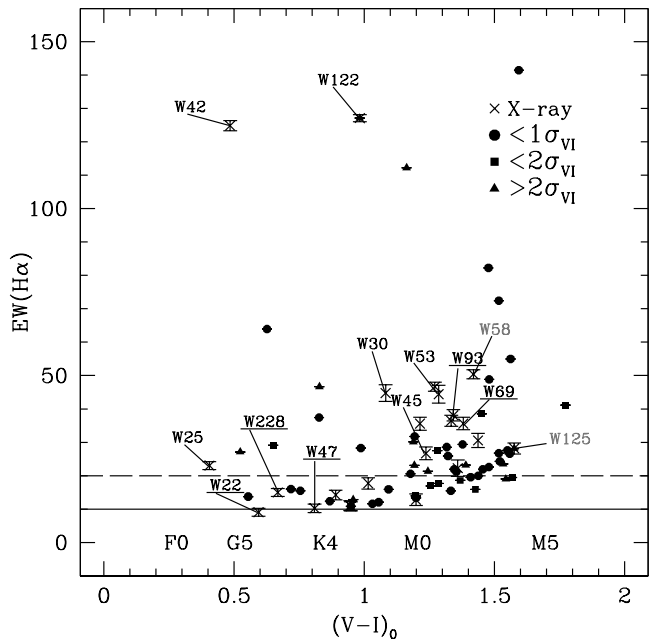} 
\caption{Distribution of EW(H$\alpha$) for stars with H$\alpha$ excess 
emission as a function of their $(V-I)_0$ colour. Symbols are the same
as in Figure\,\ref{fig_viha}. The 6 known CVs (black characters), the 5
known active binaries (ABs; underlined characters) and the 2 known
qLMXBs (grey characters) are explicitly labelled in the CMD. Spectral
types are also indicated. Note that error-bars are not shown because
they are comparable to or smaller than the symbols.} 
\label{fig_ew} 
\end{figure}

In Figure~\ref{fig_ew}, we show the distribution of the measured
EW(H$\alpha$) for all stars with H$\alpha$ excess emission detected in
47\,Tuc at 70\arcsec from the cluster centre, as a function of their $(V-I)_0$ colour and spectral type (the
latter has been computed using the models of Bessel et al. 1998). The
symbols are the same as in Figure\,\ref{fig_viha}.  It is important to
notice that all the stars selected as bona-fide H$\alpha$ excess
emitters show an EW(H$\alpha$)$\gtrsim10\AA$. This is a consequence of
the  selection criteria and of the quality of the data. In fact, because
of the photometric quality of  the data-set, only stars with an
H$\alpha$ excess with a EW(H$\alpha$)$\gtrsim10\AA$ fulfil the
requirement that $\Delta H\alpha\geqq5\times\sigma_{VH\alpha}$.

In the following section we investigate the
possibility that such a strong emission would come from pure CA.

\section{Constraints on stellar chromospheric activity}
\label{sec_ch}

The H$\alpha$ spectral line has been frequently used together with the
CaII K line as a diagnostic of CA~\citep[e.g.][]{pas91}. 
As an example, ~\citet[][]{kaf06} published a comparison of the
H$\alpha$ strengths with $V-I$ photometric colours  of eight clusters
and detected a dependence on cluster age of the  location along the MS
for the onset of CA, confirming that the  colour at which activity
becomes important is approximately a linear function of the logarithm of
the age~\citep[see also][]{haw00}. ~\citet[][]{zh11}Ê demonstrate that,
in general, CA decays with age from 50  Myr to at least 8 Gyr for stars
with spectral types later than K5 and ~\citet[][]{ly05} confirmed that
the CA of dwarfs decreases with age and remains flat in a range of ages
from  2 Gyr up to 11 Gyr (see their figure 10). It is important to
notice that the same authors suggest that, at a given age, CA  of dwarfs
increases with metallicity while, when measured through the H$\alpha$
emission, it appears to be insensitive to the activity cycle and
rotational modulation.

Furthermore, as shown by~\citet[][]{kaf06} (see their Figures 13 -- 15), 
it appears that no H$\alpha$ emission with EW
$\gtrsim 10$\,\AA\ is measured from CA for stars of spectral type K7 or earlier.
The work of~\citet[][]{kaf06} indicates that very few stars show H$\alpha$ emission with EW
$\gtrsim 10$\,\AA\ at later spectral types.~\citet[][]{sch07} show that
the CA of late-M (later than M7) dwarfs and L dwarfs can be measured as
an H$\alpha$ emission of EW(H$\alpha$)$> 10$\,\AA. Notice that these
spectral classes are not covered by our observations (see
Figure~\ref{fig_viha}). 

We shall notice that most of the CA studies refer to pop I stars close to solar metallicity.
47Tuc is relatively metal poor and not many studies have been performed so
far about CA in metal poor MS stars. A few active metal poor binaries have been studied by ~\citet[][]{pa94}.
{They  find that the H$\alpha$ emission, used as indication of CA, is observed in few active stars 
but always with EW(H$\alpha$)$< 10$\,\AA\,
taken here as benchmark.} In a sentence, given the low metallicity and old age
of 47Tuc, H$\alpha$ emission from CA is expected, if anything, to be smaller
than that from Pop I younger counterparts~\citep[see also][]{giz02}.

The stars that we selected as bona-fide H$\alpha$ emitters fall in the F0 to M5 spectral
range and all show EW larger than  $\sim 10$\,\AA. It is interesting to note that most of the mentioned
authors use a cutoff of $1.0$\,\AA\ as the minimum EW at which one can unambiguously detect
emission in an M dwarf.
Coupling the intensity of the EW(H$\alpha$) measured in this paper with the results
from the literature previously described suggests that
the H$\alpha$ excess emission that we are sampling cannot come from pure CA.

In general, the analysis above allows us to conclude that for most objects
with H$\alpha$ excess  in Figure\,\ref{fig_ew} (see also
Table~\ref{tab_emett}), and certainly for all those with EW(H$\alpha) >
20$\,\AA, the source of H$\alpha$ excess emission is not CA. These
objects could, however, be interacting binaries, in which one star fills
its Roche Lobe and matter is being transferred to the companion. We will
investigate this hypothesis in more detail in Section~\ref{sec_cmd}, but
we first look at the optical properties of known interacting binaries in
the field covered by our observations.

\section{Optical counterparts to known X-ray binaries} 
\label{sec_x}

As mentioned in the introduction, H05 have compiled a catalogue of 300
X-ray sources lying in the central regions of 47\,Tuc. The catalogue  
provides precise absolute coordinates for a number of CVs, ABs and other
exotica, but a large number of stars in that catalogue still remain
unclassified. Proving that some of these stars have an optical
counterpart showing H$\alpha$ excess emission would offer new
constraints on their nature, since recombination lines are associated
with the mass accretion process in CVs and interacting binaries in
general (see e.g. Warner 1995).

We, therefore, searched for counterparts to the H05 X-ray sources in our photometric catalogue, 
looking specifically for objects with H$\alpha$
excess emission (defined as in Section~\ref{sec_ha}) falling within $0\farcs5$ of the nominal position of
the X-ray source, which already take into account bore-sight correction.
This is a reasonable radius of tolerance given the
astrometric accuracy of the H05 catalogue and the uncertainties on our
astrometric solution ($\sim0\farcs3$).  

Most of the X-ray sources detected by H05 populate the central regions
of the cluster, where the high stellar density is more likely to favour
encounters and the formation of stellar exotica \citep[e.g.][]{kn08}. A
total of 204 X-ray sources fall inside our FoV, but 181 of them are
located within $70\arcsec$ of the cluster centre, where crowding is
severe in our observations, particularly in the V and I bands. For this
reason, we had so far excluded these regions from our analysis (see
Section\,\ref{sec_ha}), in order to minimise any systematic effects due
to photometric incompleteness. However, our goal here is to find as many counterparts
as possible to X-ray sources and to characterise their properties, so photometric incompleteness   
is no longer an issue.

Our search for counterparts to the X-ray sources over the entire FoV
revealed a total of 23 candidate matches  with H$\alpha$ excess
emission. Inside $70\arcsec$ of the cluster centre, where the number of bona-fide
H$\alpha$ excess emitters is $\sim1000$, there are 181 X-ray
sources, of which 20 have an optical counterpart with H$\alpha$ excess
emission, or $\sim 10$\,\%.
Outside of $70\arcsec$ there are 23 X-ray sources, of which 3 have an
optical counterpart, or $\sim 13$\,\%. All matches are unique, except
for one case in which there are two H$\alpha$ excess sources inside the
X-ray error circle, with rather similar $V$ and $I$ magnitudes.
Note that the relative low fraction ($\sim 10$\,\%) of matching optical
counterparts with H$\alpha$ excess is a consequence of the detection
limit of 10\,\AA\ on the value of EW(H$\alpha$), mostly imposed by the quality of
the adopted data-set (see also the discussion in Section~\ref{sec_ha}).
Nevertheless, as shown in Section~\ref{sec_ch}, this limit allows us
to exclude objects whose H$\alpha$ emission is dominated by CA. It is
therefore possible that a larger fraction of the X-ray sources in the
catalogue of H05 have an optical counterpart with H$\alpha$ emission
lines and EW(H$\alpha$)$< 10$\,\AA\,(at least at the time of our observations).
 
It is important to consider that 
the luminosity of CVs is affected by long (days) and short (hours) 
term variability~\citep[][]{ver99}. 
Interestingly, the X-ray, optical and ultraviolet flux variation during outburst and
superoutburts are often anticorrelated~\citep[][]{we03}. Moreover,~\citet[][]{fen09}
found an anticorrelation between X-ray continuum luminosity and the EW(H$\alpha$) emission 
line in X-ray binary systems. Therefore, not all objects detected as X-ray 
sources in the catalogue of H05 need to show H$\alpha$ excess
emission in the observations used in this work.

All the X-ray sources with an optical counterpart are indicated with a
cross in Figures\,\ref{fig_viha}, \ref{fig_ew} and \ref{fig_cmd}, and
their coordinates, magnitudes, {distance from the nominal position of the X-ray emitter}
and EW(H$\alpha$) values are listed in
Table\,\ref{tab_x}. Six of these objects are classified by H05 as CVs (W25,
W30, W42, W45, W53 and W122), five as ABs (W22, W47, W69, W93, W228),
two as qLMXBs (W58 and W125), while the other ten are unclassified X-ray
sources (W78, W80, W81, W145, W255, W305, W310, W312, W319 and W328).

\begin{figure}
\centering
\includegraphics[width=5cm]{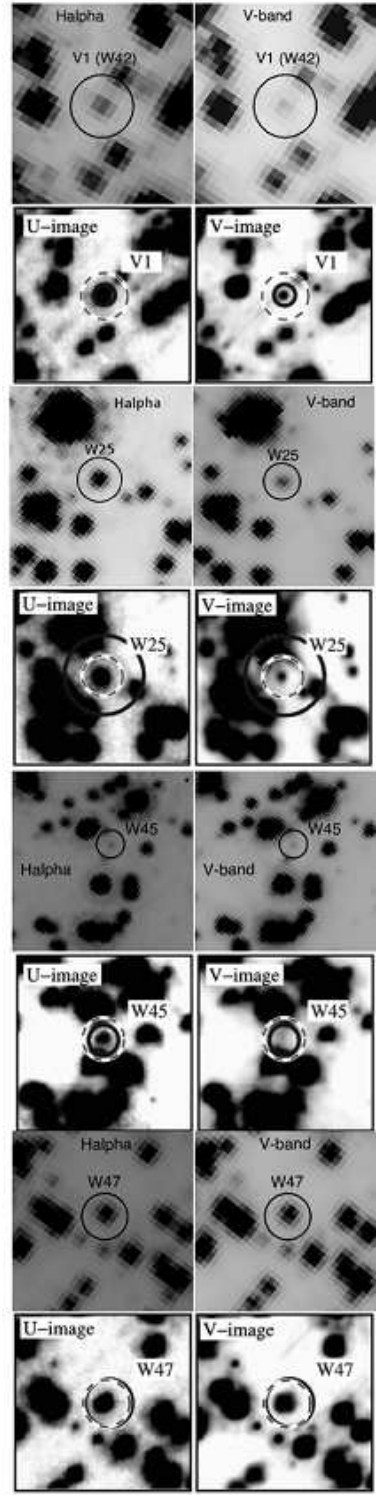}
\caption{Comparison of HST finding charts for 4 optical
identifications in common with~\citet[][]{ed03a}. For each star, the V
and H$\alpha$ finding charts from the present work (labeled V-band and Halpha,
respectively) are compared with F555W- and F336W-band images (V-band
and U-band, respectively) acquired with the WFPC2. The $0\farcs2$
solid circles show the location of the H$\alpha$ excess emitters on our
ACS images. The solid circles on the U and V WFPC2 frames show the
$4\,\sigma$  uncertainty on the X-ray positions, while the dashed circles
show the candidate optical counterparts.}
\label{fig_cvfc}
\end{figure}

In Figure~\ref{fig_cvfc} we show the finding charts from our ACS images
in the V and H$\alpha$ bands (labelled V-band and Halpha, respectively),
for three CVs namely V1 (W42), W25, W45 and the AB W47.  A $0\farcs2$
solid circle shows the location of the H$\alpha$ excess emitters on the
ACS images. We compared these objects with the finding charts in the
F336W (U) and F555W (V) bands from the WFPC2 observations
of~\citet[][]{ed03a}. While these images confirm the accuracy of our astrometric solution, it is interesting to notice 
that, thanks to the higher spatial resolution of the ACS, we identify as candidate optical counterpart 
to W45 an object that was not resolved in the photometry of~\citet[][]{ed03a}.

As Figure\,\ref{fig_ew} shows, the distribution of EW(H$\alpha$) as a
function of spectral type for X-ray sources with H$\alpha$ excess agrees
well with that of other objects with H$\alpha$ excess that we identified
and that are not detected in the X-rays. It is, therefore, possible that
the two classes of sources correspond to objects of similar types. 

%
%
%

In Figure\,\ref{fig_cmd} we show the positions of the stars with 
H$\alpha$ excess emission and EW(H$\alpha$)$ > 20$\,\AA\ in the $V_0,
(V-I)_0$ CMD, using the same symbols as in  Figure\,\ref{fig_viha}. All
cluster stars lying at a distance $r>70\arcsec$ from the cluster center
are  shown in light grey while the 6 CVs, the 5 ABs and the 2 qLMXBs are
explicitly labelled. The spectral types derived from the atmospheric
models of \citet[][]{be98} are also shown.

A photometrically unresolved binary system appears as a single star with
a flux equal to the sum of the fluxes of the two components. Thus, the
MS ridge line (solid line in Figure\,\ref{fig_cmd}) in practice
indicates the location at  which the luminosity (and hence mass) of the
secondary component is negligible with respect to the primary. The
dashed line in Figure\,\ref{fig_cmd} shows the location of binary
systems where the  masses of the two stars are equal and the magnitude
of the binary system  is $ 0.752$\,mag brighter than that of the single star
alone. These boundaries allow us to constrain the location of
unresolved binaries in the cluster.

As discussed in Section\,\ref{sec_ha}, the number of excess emitters 
increases with spectral type. In Section\,\ref{sec_ch} we explored the 
possibility that this can be a sign of unusually strong CA. Even if 
evidence of increased CA in old M dwarfs in wide binary systems with a
WD are reported in the literature~\citep[see][]{sil05}, the majority of
targets with EW(H$\alpha$) are in the range between 20\,\AA\ and 100\AA\
and can not be linked to purely chromospherically active stars.

\begin{figure}
\centering
\includegraphics[width=9cm]{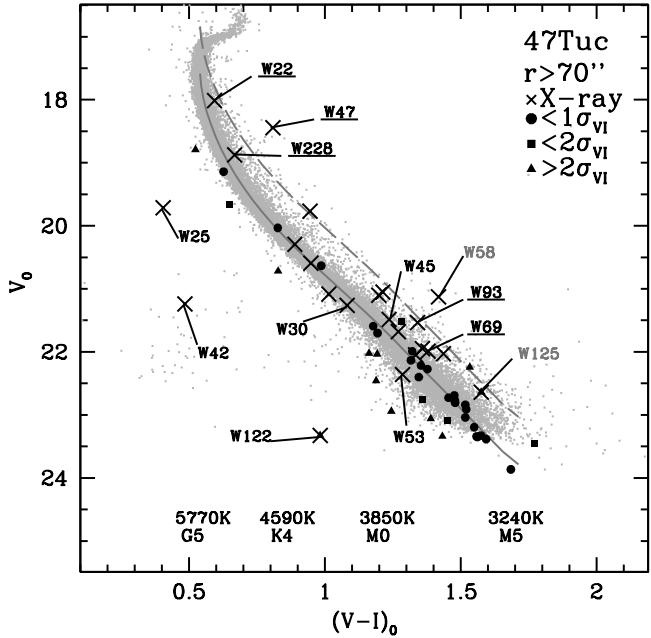}
\caption{The stars with H$\alpha$ excess emission and EW(H$\alpha>20\AA$) 
are shown as black symbols in the CMD of the cluster. 
Crosses mark the optical counterparts to X-ray sources across the whole field of view, 
whereas the grey data points and the H$\alpha$ excess sources marked with solid
circles, squares and triangles are located outside a 70\arcsec radius. The MS mean ridge line
(grey solid line) and the spectral types are also shown. The 6 known CVs (black characters), 
the 5 ABs (underlined characters) and the 2 LMXBs (grey characters) are explicitly labelled.
The MS ridge line and the location of the equal-mass binary sequence are shown with a solid 
and dashed line, respectively.}
\label{fig_cmd}
\end{figure}

An obvious alternative explanation is that these stars are compact
binaries actively interacting that were not detected in previous
studies. This conclusion seems to be supported by the fact that  in
the CMD most of these stars are located in the area between the MS ridge
line (solid line) and the line identifying the location of equal-mass
binaries (dashed line). This area is where unresolved MS binaries or BY
Draconis stars are expected~\citep[e.g.][]{al01} and where almost all
the stars showing X-ray emission and identified as active binaries by
H05 are located. Furthermore, this region may also be populated by
Algol-type binaries, which are known to undergo rapid H$\alpha$ 
emission variations~\citep[see][]{ol95}.

Even though a detailed discussion of the nature of individual
objects goes beyond the purpose and scope of the present work, sources
W22, W228, W47, W93, and W69 deserve some discussion. W47 is classified
by~\citet[][]{al01} as a semidetached W~UMa binary with a period of
$0.5$ days. Hence, a mass transfer stream, and a H$\alpha$ excess
emission, should be expected in this case. On the other hand, W\,228 is
classified as a contact W ~UMa binary,  W\,93 as a 4-day eclipsing
binary, W\,69 as a $3.1$-day BY~Dra binary, while W\,22 is a $2.5$-day
binary; none of these are likely to be engaged in mass transfer.
Nevertheless the H$\alpha$ excess emission that we find makes these objects
candidate systems actively undergoing mass transfer. 


It is interesting to note that the objects mentioned above fall on the WF detectors of the WFPC2 
camera in the observations used by Albrow et al. (2001), Edmonds et al. (2003a) and H05. These 
detectors offer a plate-scale  of $0.1\,\arcsec$\,px$^{-1}$, i.e. half of the resolution allowed by the 
ACS camera used in this work.

\begin{figure}
\centering
\includegraphics[width=8cm]{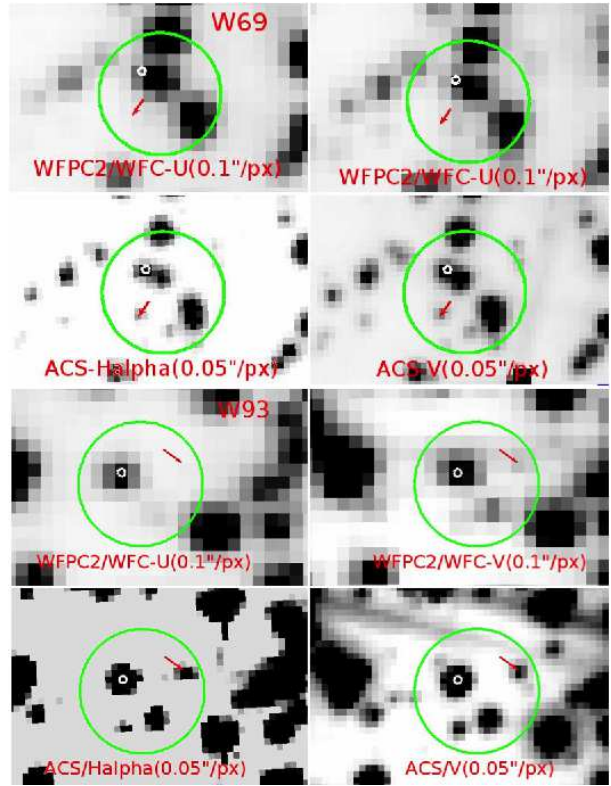}
\caption{Comparison of the FCs for the optical identifications of two X-ray objects. A large circles 
(green in the electronic version) of 0\farcs5 radius around the nominal position of the X-ray candidate are 
shown, the small white circles show the published positions of the counterparts, while the small arrow (red in the
electronic version) indicates the location of the H$\alpha$ excess emitters.}
\label{fig_cvfc}
\end{figure}

For example, as shown in Figure~\ref{fig_cvfc} by visually comparing the two data-sets we have realized 
that inside a 0\farcs5 circle around the X-ray for W69 and W93 the objects with H$\alpha$ excess emission (red arrows
in the figure) are not the same candidate optical counterparts proposed by the cited authors (white circles). Indeed, while these 
objects with H$\alpha$ excess are clearly resolved in the ACS images, they are either blended 
with nearby stars or too faint to be detected in the WFPC2 images.

We performed a test in order to quantify the probability to obtain false matching from the comparison of 
the X-ray source catalogue and the list of H$\alpha$ excess emitters. The test simply consists in applying 
an offset of few arc-seconds to the coordinates of the X-ray sources. We performed the search for optical 
counterpart among the list of H$\alpha$ emitters following the same strategy described in Section 5. We 
first performed this test for the stars outside of $70\arcsec$ from the cluster centre, where there are 23 
X-ray sources and 59 sources with H$\alpha$ excess
emis sion in our catalogue. We offset the catalogue of X-ray stars by $1\arcsec$, $2\arcsec$ and 
$3\arcsec$ and we did not find any match within $0\farcs5$. Thus, outside of $70\arcsec$ we 
expect no false matches. We then repeated this test for stars within $70\arcsec$ of the cluster
centre, where the density of sources is higher and there are 180
X-ray sources and about 1000 stars with H$\alpha$ excess emission at
the $5\,\sigma$ level. Given the higher density of objects in these regions, compared with the 
outskirts of the cluster, if we offset one of the catalogues we do find some false matches. For offsets ranging
from $1\arcsec$ to $5\arcsec$ we find typically 7 matches, of which however only typically 3 within 
a $0\farcs3$ search radius. 
We therefore conclude that, in the worst scenario, 
7 false matches should be expected among the 20 matches.

Non-periodic strong variability would also offer an explanation to the
unusual location of stars W\,47 (AB) and W\,58 (qLX). W\,47 shows clear
signs of variability in the Chandra observations, as reported by H05
(see their Figure\,6). The same authors identify W\,47 with an optical
variable, corresponding to object number 8 in the list
of~\citet[][]{ed96} and PC1--V08 in that of~\citet[][]{al01}. Star W\,58
is also classified as variable by H05 and ~\citet[][]{ed03a}.  While the
finding charts shown in Figure~\ref{fig_cvfc} demonstrate the good match
in our source identification, the location of these two stars on the CMD
is not in agreement with the one reported by \citet[][]{ed96} and
\citet[][]{al01}. However, as a consistency check of our photometric
measurements, we have looked for these objects in the photometric
catalogue of the same region provided as part of the ACS Globular
Cluster Treasury programme ~\citep[][]{sa07} \footnote{The catalogue is
available online at
http://www.astro.ufl.edu/\~ata/public\_hstgc/databases.html.}. The
magnitudes that we measured in the V (F606W) and I (F814W) bands are
fully consistent with those of the ACS Globular Cluster Treasury
programme. We conclude, therefore, that  our photometry is correct and
that these two sources are likely to be affected by  episodes of large
variability (we show in Table~\ref{tab_x} the timescale of the
variability at the 99\,\% and $99.9\,\%$ confidence levels as derived by
H05).

Stars with H$\alpha$ excess in the colour range $1 < (V-I)_0 < 1.5$ and
lying away from the MS ridge line by more than twice their colour
uncertainty $\sigma_{VI}$ represent a rather interesting sub-class of
objects. These stars, indicated by solid triangles in
Figure\,\ref{fig_cmd}, have EW(H$\alpha$) in the range between $\sim
20$\,\AA\ and $\sim 130$\,\AA\, suggesting that they are accretors.
These stars systematically populate the bluest side of the MS, far from
the location of binary stars. Since a CV showing X-ray and H$\alpha$
emission, namely W53, is located in the same region of the CMD, we
conclude that these objects too are indeed most likely candidate CVs.

\section{Discussion and conclusions}
\label{sec_cmd}

The CMD is often used to constrain the nature of X-ray sources in  a
GC~\citep[e.g.][]{coh10}. In a study devoted to the detection of CVs in
the GC NGC~6752, ~\citet[][hereafter B96]{ba96} identified two CVs from
their H$\alpha$ and UV excess using observations with the WFPC2 camera
on board the HST. These authors notice that the location of these stars
in the vicinity of the MS in the $V, V-I$ CMD (the same location as
occupied by our candidates; see also their Figure\,3) is not typical for
CVs. B96 suggest that the red colours of the counterpart may indicate
that there is a significant contribution from a low-mass secondary to
the optical emission from the source. 

The EW(H$\alpha$) values that we have measured for these objects strongly
support the hypothesis that they are bona-fide CVs. \citet[][]{pe08}
used observations from the AAO/UKST SuperCOSMOS H$\alpha$
Survey~\citep[SHS;][]{pa05} to identify CVs through their H$\alpha$
emission. They found 16 CVs with spectroscopically measured EW(H$\alpha$)
in the range between $\sim 20$\,\AA\ and $\sim 130$\,\AA, in excellent
agreement with the values we obtained for the 6 CVs with X-ray and
H$\alpha$ emission (see Table~\ref{tab_x}) and with the measured
EW(H$\alpha$) of the candidate CVs (see Figures~\ref{fig_viha}
and~\ref{fig_ew}). 

This finding opens up an interesting new scenario. Theoretical models
for CV formation in GCs predict that a cluster like 47 Tuc should harbor
$\sim 200$ CVs, most of which should have formed
dynamically~\citep[][]{di94,iv06,po06}. About 50\,\% of these objects
are expected to reside in the cluster core. The combined
observational effort described by H05 and~ \citet[][]{kn08}, including
all the detections obtained at X-ray, optical and UV wavelengths,
together with spectroscopy from the HST, bring the total number of
candidate CVs detected so far in 47 Tuc to $\sim 100$ CVs.  Our
observations suggest that this number is likely to grow and might
approach the level predicted by theoretical models.

Let us assume that the density distribution of these known CV candidates
is representative of the radial distribution of the entire population of
CVs in 47 Tuc. A total of 15 of the X-ray CVs are located inside the
core radius~\citep[$\sim$ 21\arcsec;][]{map04}, while only 2 are found
outside of 70\arcsec. Our selection based on H$\alpha$ excess emission
in the region outside of 70\arcsec, reveals 12 objects with
EW(H$\alpha$)$>20\AA$ and showing a $(V-I)_0$ colour at least
$>1\sigma_{VI}$ bluer than the MS ridge line (solid black squares and
triangles in Figure~\ref{fig_cmd}). We include also the X-ray source
W\,122, classified as a CV by H05. Under the assumption that the ratio of
H$\alpha$ and X-ray emitters is constant and of order 6 everywhere, we
should expect to detect $\sim 90$ CVs with H$\alpha$ emission inside the
core radius, a number in full agreement with the theoretical
expectations. Therefore, by allowing an efficient detection of candidate
CVs, our method is able to reconcile theoretical expectation with
observations. 

It is appropriate to point out that the validity of the
observational approach presented in this paper was already exploited
by~\citet[][hereafter W06]{wit06}. These authors used observations from
the Isaac Newton Telescope Photometric H$\alpha$  Survey of the northern
galactic plane (IPHAS) to study the H$\alpha$ emission of 71 known CVs. 
The selection strategy presented in our paper is quite similar to the
one adopted in W06, with the difference that we use the V band instead
of the $r'$ filter to estimate the stellar continuum below the H$\alpha$
line. The advantage of our approach are described in
Section~\ref{sec_ha}. Nevertheless, it is comforting that W06 report a
70\% recovery rate of known CVs as H$\alpha$ emitters and that they do
not find statistically significant dependence of the recovery rate of
CVs  via their H$\alpha$ emission on the actual CV type. 

An important result of this study is that chromospheric activity alone
is not sufficient to produce H$\alpha$ excess emission with
EW(H$\alpha$)$> 20\,\AA$ and that the origin of such a strong emission
has to be associated with mass accretion. This is crucial for the
identification and study of not only interacting binaries but also of T
Tauri stars in star forming regions. Our results show that the method
developed by DM10 to derive EW(H$\alpha$) from a suitable combination of
photometry in the H$\alpha$, V and I bands can effectively distinguish
stars undergoing mass accretion from chromospherically active objects. 

\section*{Acknowledgments}

We are very grateful to an anonymous referee for the constructive
comments and useful suggestions that have  helped us to improve the
overall presentation of the scientific results. The authors thank Paolo
Montegriffo for his help with the CataXcorr software package. The
research leading to these results has received funding from the European
Community's Seventh  Framework Programme (/FP7/2007-2013/) under grant
agreement No 229517.


\begin{center}
\onecolumn
\begin{longtable}{c c c c c c c}
\caption{Candidate H$\alpha$ excess emitters.
\label{tab_emett}} \\

\hline
Name & RA & Dec & V & I & H$\alpha$ & EW(H$\alpha$) [\AA]) \\  \hline

\endfirsthead

\hline
Name & RA & Dec & V & I & H$\alpha$ & EW(H$\alpha$) [\AA]) \\  \hline
\endhead

\hline Continued on next page \\ \hline
\endfoot

\hline 
\endlastfoot
     1&  00:24:06.82 &  -72:06:09.87  &  18.546& 17.952& 18.010 &  13.74 $\pm$  1.60	  \\
     2&  00:24:17.71 &  -72:05:33.38  &  18.907& 18.344& 18.233 &  27.12 $\pm$  2.13	  \\
     3&  00:24:12.44 &  -72:06:11.22  &  19.255& 18.589& 18.196 &  63.91 $\pm$  1.41	  \\
     4&  00:24:18.35 &  -72:06:14.64  &  19.644& 18.850& 18.977 &  15.47 $\pm$  1.41	  \\
     5&  00:24:17.50 &  -72:05:35.26  &  19.673& 18.915& 19.021 &  15.99 $\pm$  1.43	  \\
     6&  00:23:49.91 &  -72:04:11.30  &  19.770& 19.081& 19.011 &  29.15 $\pm$  2.03	  \\
     7&  00:24:15.86 &  -72:05:56.50  &  20.147& 19.281& 19.203 &  37.43 $\pm$  2.29	  \\
     8&  00:24:20.37 &  -72:05:35.68  &  20.299& 19.392& 19.607 &  12.34 $\pm$  1.50	  \\
     9&  00:23:54.62 &  -72:06:07.68  &  20.709& 19.721& 19.994 &  10.83 $\pm$  1.34	  \\
    10&  00:24:21.07 &  -72:05:28.60  &  20.748& 19.722& 19.813 &  28.26 $\pm$  1.56	  \\
    11&  00:23:58.48 &  -72:06:30.11  &  20.833& 19.966& 19.803 &  46.63 $\pm$  2.21	  \\
    12&  00:23:45.94 &  -72:03:46.16  &  20.944& 19.873& 20.178 &  11.57 $\pm$  1.50	  \\
    13&  00:23:46.94 &  -72:05:27.25  &  21.139& 20.044& 20.355 &  12.06 $\pm$  1.50	 \\
    14&  00:23:43.70 &  -72:04:42.17  &  21.248& 20.115& 20.398 &  15.93 $\pm$  1.67	 \\
    15&  00:24:21.43 &  -72:04:15.97  &  21.353& 20.356& 20.608 &  12.86 $\pm$  1.86	  \\
    16&  00:24:18.93 &  -72:04:17.59  &  21.517& 20.279& 20.647 &  13.35 $\pm$  1.55	  \\
    17&  00:23:55.54 &  -72:06:29.16  &  21.627& 20.307& 20.556 &  27.52 $\pm$  2.64	  \\
    18&  00:23:52.64 &  -72:03:59.05  &  21.708& 20.491& 20.763 &  20.56 $\pm$  1.81	  \\
    19&  00:23:51.79 &  -72:06:03.40  &  21.816& 20.583& 20.743 &  31.75 $\pm$  2.52	  \\
    20&  00:24:19.70 &  -72:04:09.27  &  21.997& 20.760& 21.120 &  13.96 $\pm$  1.77	  \\
    21&  00:23:45.70 &  -72:04:27.72  &  22.106& 20.745& 21.033 &  25.86 $\pm$  2.03	  \\
    22&  00:23:49.52 &  -72:05:38.62  &  22.136& 20.934& 20.468 & 112.23 $\pm$  1.10	  \\
    23&  00:23:52.34 &  -72:05:43.39  &  22.150& 20.918& 21.169 &  23.12 $\pm$  2.16	  \\
    24&  00:23:55.61 &  -72:06:00.78  &  22.212& 20.840& 21.252 &  15.46 $\pm$  1.83	  \\
    25&  00:23:54.57 &  -72:05:49.39  &  22.227& 20.933& 21.287 &  16.92 $\pm$  1.72	  \\
    26&  00:24:20.41 &  -72:04:59.93  &  22.247& 20.664& 21.141 &  19.06 $\pm$  1.73	  \\
    27&  00:23:47.23 &  -72:05:34.88  &  22.248& 20.892& 21.149 &  28.54 $\pm$  2.22	  \\
    28&  00:23:47.71 &  -72:04:49.64  &  22.310& 20.986& 21.347 &  17.65 $\pm$  2.02	  \\
    29&  00:23:46.51 &  -72:05:31.92  &  22.331& 20.938& 21.294 &  21.20 $\pm$  2.11	  \\
    30&  00:23:46.18 &  -72:03:47.22  &  22.357& 20.784& 21.205 &  23.59 $\pm$  2.76	  \\
    31&  00:23:51.23 &  -72:05:23.88  &  22.387& 20.970& 21.250 &  29.37 $\pm$  1.96	  \\
    32&  00:24:01.62 &  -72:06:02.79  &  22.448& 20.982& 21.437 &  15.92 $\pm$  1.79	  \\
    33&  00:23:53.19 &  -72:05:35.48  &  22.515& 21.130& 21.474 &  21.90 $\pm$  1.71	  \\
    34&  00:23:52.18 &  -72:03:48.37  &  22.571& 21.343& 21.517 &  30.14 $\pm$  1.94	  \\
    35&  00:23:51.85 &  -72:05:28.97  &  22.620& 21.172& 21.575 &  19.57 $\pm$  2.20	  \\
    36&  00:23:53.97 &  -72:03:50.22  &  22.754& 21.139& 21.531 &  28.18 $\pm$  1.65	  \\
    37&  00:23:43.75 &  -72:04:32.68  &  22.804& 21.288& 21.173 &  82.34 $\pm$  1.21	  \\
    38&  00:24:23.96 &  -72:05:13.79  &  22.843& 21.348& 21.748 &  21.96 $\pm$  1.83	  \\
    39&  00:23:48.33 &  -72:05:16.16  &  22.866& 21.467& 20.345 & 301.35 $\pm$  0.50	  \\
    40&  00:24:06.40 &  -72:06:26.63  &  22.882& 21.474& 21.868 &  18.56 $\pm$  2.12	  \\
    41&  00:23:47.93 &  -72:04:20.71  &  22.884& 21.366& 21.770 &  22.65 $\pm$  2.11	  \\
    42&  00:24:01.21 &  -72:05:58.79  &  22.907& 21.430& 21.843 &  20.00 $\pm$  2.10	  \\
    43&  00:23:50.11 &  -72:04:09.17  &  22.920& 21.401& 21.547 &  48.89 $\pm$  1.68	  \\
    44&  00:24:01.58 &  -72:06:03.14  &  22.923& 21.316& 21.799 &  19.55 $\pm$  2.05	  \\
    45&  00:24:04.39 &  -72:06:01.84  &  22.954& 21.398& 21.776 &  26.78 $\pm$  2.09	  \\
    46&  00:23:58.23 &  -72:06:22.33  &  23.025& 21.465& 21.871 &  24.37 $\pm$  2.29	  \\
    47&  00:24:19.86 &  -72:05:50.88  &  23.057& 21.773& 22.071 &  21.33 $\pm$  2.30	  \\
    48&  00:23:52.06 &  -72:05:52.25  &  23.154& 21.598& 21.573 &  72.52 $\pm$  1.44	  \\
    49&  00:24:02.22 &  -72:06:12.22  &  23.174& 21.745& 22.098 &  23.14 $\pm$  2.36	  \\
    50&  00:23:47.83 &  -72:05:43.70  &  23.202& 21.711& 21.937 &  38.60 $\pm$  2.26	  \\
    51&  00:24:19.30 &  -72:05:41.18  &  23.310& 21.721& 22.106 &  27.65 $\pm$  2.30	  \\
    52&  00:24:03.84 &  -72:06:21.71  &  23.441& 22.419& 21.779 & 127.08 $\pm$  1.08	  \\
    53&  00:23:55.24 &  -72:05:44.63  &  23.442& 21.828& 20.658 & 359.66 $\pm$  0.45	  \\
    54&  00:24:22.34 &  -72:04:11.02  &  23.453& 21.981& 20.725 & 365.81 $\pm$  0.48	  \\
    55&  00:23:55.32 &  -72:05:53.33  &  23.455& 21.857& 22.257 &  26.65 $\pm$  2.33	  \\
    56&  00:23:50.78 &  -72:04:07.57  &  23.467& 21.866& 22.000 &  55.04 $\pm$  1.63	  \\
    57&  00:24:12.62 &  -72:06:08.16  &  23.499& 21.866& 21.462 & 141.46 $\pm$  0.93	  \\
    58&  00:23:51.60 &  -72:05:51.52  &  23.574& 21.762& 22.122 &  40.24 $\pm$  2.23	  \\
    59&  00:24:21.68 &  -72:05:46.24  &  23.979& 22.255& 20.707 & 569.99 $\pm$  0.36	  \\
\hline
\end{longtable}
\end{center}

\vspace{10cm}

\begin{table}
\begin{center}
\caption{
Objects with X-ray emission from the catalogue of H05 showing H$\alpha$ excess emission. We use the names of the sources as given
by H05 in their Table\,2.
\label{tab_x}}
\begin{tabular}{c c c c c c c c c c}
\hline
Name & RA & Dec & V & I & H$\alpha$ & EW(H$\alpha$) [\AA] & distance [arcsec] & classification$^{\ast}$\\
\hline
W25  	&   00:24:07.131 & -72:05:45.85    & 19.830  &  19.386  & 19.256   &   23.05$\pm$1.20  &   0.07      &	     CV (D)	\\
W45  	&   00:24:03.762 & -72:04:22.82    & 21.601  &  20.325  & 20.560   &   26.69$\pm$1.92  &    0.16      &        CV  (DH) \\
W30  	&   00:24:05.991 & -72:04:56.24    & 21.378  &  20.257  & 20.235   &   44.75$\pm$2.47  &     0.06     &        CV  (DY) \\
W42  	&   00:24:04.244 & -72:04:58.11    & 21.355  &  20.831  & 19.970   &  124.81$\pm$1.53  &    0.09      &        CV  (DHY)  \\
W122 	&   00:24:03.840 & -72:06:21.71    & 23.441  &  22.419  & 21.779   &  127.09$\pm$1.08  &  0.10	     &        CV (D)	   \\
W53  	&   00:24:02.509 & -72:05:10.80    & 22.474  &  21.149  & 21.235   &   44.42$\pm$2.67   &   0.46      &        CV  (Y) \\
W22  	&   00:24:07.822 & -72:05:24.44    & 18.129  &  17.496  & 17.633   &    9.04$\pm$1.19   &    0.11     &        AB   \\
W47  	&   00:24:03.452 & -72:05:05.35    & 18.558  &  17.710  & 17.925   &   10.22$\pm$1.24   &   0.07      &        AB  (DHY)  \\
W69  	&   00:24:12.704 & -72:04:22.72    & 22.105  &  20.686  & 20.904   &   35.61$\pm$1.87	&  0.46      &        AB   \\
W228 	&   00:24:18.596 & -72:04:55.34    & 18.989  &  18.282  & 18.379   &   14.96$\pm$1.16   &    0.25     &        AB   \\
W93  	&   00:24:12.073 & -72:05:07.47    & 21.650  &  20.269  & 20.443   &   38.11$\pm$1.68   &    0.38     &        AB   \\
W58  	&   00:24:00.946 & -72:04:53.31    & 21.243  &  19.785  & 19.887   &   50.38$\pm$1.38   &    0.11    &	     qLX  (DHY)\\
W125 	&   00:23:53.975 & -72:03:50.22    & 22.754  &  21.139  & 21.531   &   28.18$\pm$1.65   &   0.15      &        qLX (D?H) \\
W319 	&   00:23:58.849 & -72:04:47.38    & 22.061  &  20.663  & 21.010   &   22.24$\pm$2.44   &  0.37       &    (H?) \\
W80  	&   00:24:02.540 & -72:04:41.04    & 19.881  &  18.896  & 19.165   &   11.03$\pm$1.49   &  0.49       &        \\
W312 	&   00:24:00.516 & -72:05:17.26    & 20.407  &  19.478  & 19.680   &   14.24$\pm$1.43  &   0.41       &         \\	  
W145 	&   00:23:54.625 & -72:06:07.68    & 20.709  &  19.721  & 19.994   &   10.83$\pm$1.34	&   0.24     &         \\	  
W255 	&   00:24:10.566 & -72:04:43.91    & 21.164  &  19.911  & 20.043   &   35.59$\pm$1.94  &    0.30      &         \\	  
W78  	&   00:24:03.687 & -72:05:22.26    & 21.197  &  20.143  & 20.365   &   17.72$\pm$1.70  &    0.30      &         \\	  
W310 	&   00:24:00.819 & -72:05:42.77    & 21.216  &  19.978  & 20.353   &   12.79$\pm$1.75  &   0.30       &         \\	  
W81  	&   00:24:01.799 & -72:05:12.78    & 21.794  &  20.484  & 20.542   &   46.66$\pm$1.36  &    0.47      &         \\	  
W328 	&   00:23:56.328 & -72:04:37.84    & 22.144  &  20.668  & 20.966   &   30.57$\pm$2.13  &  0.47	     &         \\	  
W305 	&   00:24:02.003 & -72:04:25.47    & 22.162  &  20.789  & 20.975   &   36.44$\pm$1.68  &  0.19	     &         \\	  
\hline
\end{tabular}
\end{center}
\tablecomments{$\ast$ Source classification taken from H05; CV =
Cataclysmic variables; qLX = low-mass X-ray binaries containing accreting
neutron stars; AB = active binaries. The variability timescale of each
star and the associated confidence level is indicated in parentheses as
follow: D, H and Y for 99.9\,\% variability confidence on day, hours and
years timescale, respectively. D? and H? indicate a 99\,\% variability
confidence level on timescales of days and hours, respectively.}
\end{table}

\bsp

\label{lastpage}


\begin{thebibliography}{}

\bibitem[Albrow et al.(2001)]{al01} Albrow, M.~D., 
Gilliland, R.~L., Brown, T.~M., et al.\ 2001, ApJ, 559, 1060 

\bibitem[Anderson et al.(2009)]{an09} Anderson, J., Piotto, G., King,
I.~R., Bedin, L.~R., \& Guhathakurta, P.\ 2009, ApJL, 697, L58 

\bibitem[Anderson \& Bedin(2010)]{an10} Anderson, J., \& Bedin, L.~R.\ 2010, PASP, 122, 1035 

\bibitem[Appenzeller \& Mundt(1989)]{ap89} Appenzeller, I., \& Mundt,
R.\ 1989, A\&ARv, 1, 291

\bibitem[Bailyn(1995)]{ba95} Bailyn, C.~D.\ 1995, ARA\&A, 33, 133 

\bibitem[Bailyn et al.(1996)]{ba96} Bailyn, C.~D., Rubenstein, E.~P.,
Slavin, S.~D., et al.\ 1996, ApJL, 473, L31 

\bibitem[Baraffe et al.(1997)]{ba97} Baraffe, I., Chabrier, G., Allard, F., \& Hauschildt, P.~H.\ 1997, A\&A, 327, 1054 

\bibitem[Bassa et al.(2008)]{ba08} Bassa, C.~G., Pooley, D., Verbunt, F., et al.\ 2008, A\&A, 488, 921 

\bibitem[Beccari et al.(2010)]{be10} Beccari, G., et al.\ 2010, ApJ,
720, 1108 
 
\bibitem[Bertin \& Arnouts(1996)]{be96} Bertin, E., \& Arnouts, S.\ 1996, A\&AS, 117, 393 

\bibitem[Bertout(1989)]{ber89} Bertout, C.\ 1989, ARA\&A, 27, 351 

\bibitem[Bessell et al.(1998)]{be98} Bessell, M.~S., Castelli, F., \&
Plez, B.\ 1998, A\&A, 333, 231 

\bibitem[Camilo et al.(2000)]{ca00} Camilo, F., Lorimer, 
D.~R., Freire, P., Lyne, A.~G., \& Manchester, R.~N.\ 2000, ApJ, 535, 975 

\bibitem[Cardelli et al.(1989)]{car89} Cardelli, J.~A., 
Clayton, G.~C., \& Mathis, J.~S.\ 1989, ApJ, 345, 245 

\bibitem[Carretta et al.(2005)]{ca05} Carretta, E., Gratton, R.~G.,
Lucatello, S., Bragaglia, A., \& Bonifacio, P.\ 2005, A\&A, 433, 597 

\bibitem[Carretta et al.(2009)]{ca09} Carretta, E., Bragaglia, A., Gratton, R., D'Orazi, V., \& Lucatello, S.\ 2009, A\&A, 508, 695 

\bibitem[Cohn et al.(2010)]{coh10} Cohn, H.~N., Lugger, P.~M., Couch,
S.~M., et al.\ 2010, ApJ, 722, 20 


\bibitem[De Marchi et al.(2010)]{de10} De Marchi, G., Panagia, N., \&
Romaniello, M.\ 2010, ApJ, 715, 1 (DM10)

\bibitem[De Marchi et al.(2011)]{de11} De Marchi, G., Panagia, N.,
Romaniello, M., et al. 2011, ApJ, 740, 11

\bibitem[Di Stefano \& Rappaport(1994)]{di94} Di Stefano, R., \& Rappaport, S.\ 1994, ApJ, 423, 274 

\bibitem[Edmonds et al.(1996)]{ed96} Edmonds, P.~D., 
Gilliland, R.~L., Guhathakurta, P., et al.\ 1996, ApJ, 468, 241 

\bibitem[Edmonds et al.(2003a)]{ed03a} Edmonds, P.~D., Gilliland, R.~L.,
Heinke, C.~O., \& Grindlay, J.~E.\ 2003a, ApJ, 596, 1177 

\bibitem[Edmonds et al.(2003b)]{ed03b} Edmonds, P.~D., Gilliland, R.~L., Heinke, 
C.~O., \& Grindlay, J.~E.\ 2003b, ApJ, 596, 1197 

\bibitem[Fender et al.(2009)]{fen09} Fender, R.~P., Russell, D.~M., Knigge, C., et al.\ 2009, MNRAS, 393, 1608 

\bibitem[Ferraro et al.(2001)]{fe01} Ferraro, F.~R., D'Amico, N.,
Possenti, A., Mignani, R.~P., \& Paltrinieri, B.\ 2001, ApJ, 561, 337 

\bibitem[Ferraro et al.(2004)]{fe04} Ferraro, F.~R., 
Beccari, G., Rood, R.~T., et al.\ 2004, ApJ, 603, 127 

\bibitem[Gizis et al.(2002)]{giz02} Gizis, J.~E., Reid, 
I.~N., \& Hawley, S.~L.\ 2002, AJ, 123, 3356 

\bibitem[Grindlay et al.(2001)]{gr01} Grindlay, J.~E., Heinke, C.,
Edmonds, P.~D., \& Murray, S.~S.\ 2001, Science, 292, 2290 

\bibitem[Harris(1996)]{ha96} Harris, W.E. 1996, AJ, 112, 1487

\bibitem[Hawley et al.(1996)]{haw96} Hawley, S.~L., Gizis, J.~E., \&
Reid, I.~N.\ 1996, AJ, 112, 2799 

\bibitem[Hawley et al.(2000)]{haw00} Hawley, S.~L., Reid, I.~N., \&
Tourtellot, J.~G.\ 2000, Very Low-Mass Stars and Brown Dwarfs, 109 

\bibitem[Heinke et al.(2005)]{he05} Heinke, C.~O., Grindlay, J.~E.,
Edmonds, P.~D., et al.\ 2005, ApJ, 625, 796 (H05)

\bibitem[Ivanova et al.(2006)]{iv06} Ivanova, N., Heinke, C.~O., Rasio, F.~A., et al.\ 2006, MNRAS, 372, 1043 

\bibitem[Kafka \& Honeycutt(2006)]{kaf06} Kafka, S., \& Honeycutt,
R.~K.\ 2006, AJ, 132, 1517 

\bibitem[Kalirai et al.(2012)]{kal12} Kalirai, J.~S., Richer, 
H.~B., Anderson, J., et al.\ 2012, AJ, 143, 11 

\bibitem[Koenigl(1991)]{ko91} Koenigl, A.\ 1991, ApJL, 370, L39

\bibitem[Knigge et al.(2002)]{k02} Knigge, C., Zurek, D.~R., Shara,
M.~M., \& Long, K.~S.\ 2002, ApJ, 579, 752

\bibitem[Knigge et al.(2008)]{kn08} Knigge, C., Dieball, A., Ma{\'{\i}}z
Apell{\'a}niz, J., et al.\ 2008, ApJ, 683, 1006 

\bibitem[Knigge et al.(2011)]{kn11} Knigge, C., Baraffe, I., \&
Patterson, J.\ 2011, ApJS, 194, 28 

\bibitem[Kong et al.(2006)]{kon06} Kong, A.~K.~H., Bassa, C., 
Pooley, D., et al.\ 2006, ApJ, 647, 1065 

\bibitem[Linsky \& Ayres(1978)]{li78} Linsky, J.~L., \& Ayres, T.~R.\ 1978, ApJ, 220, 619 

\bibitem[Linsky(1980)]{ly80} Linsky, J.~L.\ 1980, ARA\&A, 18, 439 

\bibitem[Lu et al.(2011)]{lu11} Lu, T.-N., Kong, A.~K.~H., Verbunt, F.,
et al.\ 2011, ApJ, 736, 158 

\bibitem[Lyra \& Porto de Mello(2005)]{ly05} Lyra, W., \& Porto de Mello, G.~F.\ 2005, A\&A, 431, 329 

\bibitem[Mapelli et al.(2004)]{map04} Mapelli, M., Sigurdsson, S.,
Colpi, M., et al.\ 2004, ApJL, 605, L29 

\bibitem[Milone et al.(2011)]{mi11} Milone, A.~P., Piotto, G., Bedin,
L.~R., et al.\ 2011, arXiv:1111.0552 

\bibitem[Olson \& Etzel(1995)]{ol95} Olson, E.~C., \& Etzel, P.~B.\ 1995, AJ, 109, 1308 

\bibitem[Pace \& Pasquini(2004)]{pa04} Pace, G., \& Pasquini, L.\ 2004,
A\&A, 426, 1021

\bibitem[Pallanca et al.(2010)]{pa10} Pallanca, C., Dalessandro, E.,
Ferraro, F.~R., et al.\ 2010, ApJ, 725, 1165 

\bibitem[Pallanca et al.(2013)]{pa13} Pallanca, C., 
Dalessandro, E., Ferraro, F.~R., Lanzoni, B., 
\& Beccari, G.\ 2013, ApJ, 773, 122 


\bibitem[Parker et al.(2005)]{pa05} Parker, Q.~A., Phillipps, S.,
Pierce, M.~J., et al.\ 2005, MNRAS, 362, 689 

\bibitem[Pasquini \& Pallavicini(1991)]{pas91} Pasquini, L., \&
Pallavicini, R.\ 1991, A\&A, 251, 199 

\bibitem[Pasquini \& Lindgren(1994)]{pa94} Pasquini, L., \& Lindgren, H.\ 1994, A\&A, 283, 179 

\bibitem[Pooley et al.(2002)]{po02} Pooley, D., Lewin, 
W.~H.~G., Homer, L., et al.\ 2002, ApJ, 569, 405 

\bibitem[Pooley \& Hut(2006)]{po06} Pooley, D., \& Hut, P.\ 2006, ApJL, 646, L143 

\bibitem[Pretorius \& Knigge(2008)]{pe08} Pretorius, M.~L., \& Knigge,
C.\ 2008, MNRAS, 385, 1471 


\bibitem[Rocha-Pinto et al.(2002)]{ro02} Rocha-Pinto, H.~J., Castilho,
B.~V., \& Maciel, W.~J.\ 2002, A\&A, 384, 912 

\bibitem[Sarajedini et al.(2007)]{sa07} Sarajedini, A., 
Bedin, L.~R., Chaboyer, B., et al.\ 2007, AJ, 133, 1658 

\bibitem[Schmidt et al.(2007)]{sch07} Schmidt, S.~J., Cruz, K.~L.,
Bongiorno, B.~J., Liebert, J., \& Reid, I.~N.\ 2007, AJ, 133, 2258 

\bibitem[Sicilia-Aguilar et al.(2011)]{si11} Sicilia-Aguilar, A.,
Henning, T., Kainulainen, J., \& Roccatagliata, V.\ 2011, ApJ, 736,
137 

\bibitem[Shu et al.(1994)]{shu94} Shu, F.~H., Najita, J., Ruden, S.~P.,
\& Lizano, S.\ 1994, ApJ, 429, 797 

\bibitem[Silvestri et al.(2005)]{sil05} Silvestri, N.~M., Hawley, S.~L.,
\& Oswalt, T.~D.\ 2005, AJ, 129, 2428 

\bibitem[Sirianni et al.(2005)]{sir05} Sirianni, M., et al.\ 2005,
PASP, 117, 1049 

\bibitem[Spezzi et al.(2012)]{spe12} Spezzi, L., De Marchi,  G.,
Panagia, N., Sicilia-Aguilar, A., \& Ercolano, B.\ 2011, MNRAS, in
press, arXiv:1111.0835 

\bibitem[Stetson(1987)]{ste87} Stetson, P.~B.\ 1987, PASP, 99, 191 

\bibitem[Stetson(1994)]{ste94} Stetson, P.B., 1994, PASP, 106, 250

\bibitem[Verbunt et al.(1999)]{ver99} Verbunt, F., Wheatley, P.~J., \& Mattei, J.~A.\ 1999, A\&A, 346, 146 

\bibitem[Warner(1995)]{war95} Warner, B. 1995, Cataclysmic variable
stars, Cambridge Astrophysics Series, Cambridge: Cambridge University 
Press

\bibitem[West et al.(2004)]{we04} West, A.~A., Hawley, S.~L., Walkowicz,
L.~M., et al.\ 2004, AJ, 128, 426 

\bibitem[Wheatley et al.(2003)]{we03} Wheatley, P.~J., Mauche, C.~W., \& Mattei, J.~A.\ 2003, MNRAS, 345, 49 

\bibitem[White \& Basri(2003)]{wh03} White, R.~J., \& Basri, G.\ 2003,
ApJ, 582, 1109 

\bibitem[Williams(1983)]{wil83} Williams, G.\ 1983, ApJS, 53, 523 

\bibitem[Witham et al.(2006)]{wit06} Witham, A.~R., Knigge, 
C., G{\"a}nsicke, B.~T., et al.\ 2006, MNRAS, 369, 581 

\bibitem[Zhao et al.(2011)]{zh11} Zhao, J.~K., Oswalt, T.~D., Rudkin,
M., Zhao, G., \& Chen, Y.~Q.\ 2011, AJ, 141, 107 

\end{thebibliography}
\end{document}